\documentclass{ws-p10x7}

\def\gtrsim{\mathrel{\hbox{\rlap{\hbox{\lower4pt\hbox{$\sim$}}}\hbox{$>$}}}}

\newcommand{\beq}{\begin{equation}}
\newcommand{\eeq}{\end{equation}}
\def\beqa{\begin{eqnarray}}
\def\eeqa{\end{eqnarray}}

\newcommand{\gsim}{\gtrsim}

\begin{document}

\title{COSMOLOGY AND PARTICLE PHYSICS}  

\author{MASATAKA FUKUGITA}
\address{Institute for Cosmic Ray Research, University of Tokyo, 
Kashiwa 2778582, Japan\\
Institute for Advanced Study, Princeton, NJ 08540, USA}

\twocolumn[\maketitle\abstract{ 
The state of our understanding of cosmology is reviewed from an
astrophysical cosmologist point of view with a particular emphasis
given to recent observations and their impact. 
Discussion is then presented on the implications for particle physics.
}]





\section{Overview of astrophysical cosmology}

By 1970 astrophysicists were already reasonably confident that the
universe began as a fireball and the hot universe 
then cooled as it expanded. The 2.7K radiation and
the 25\% helium abundance are among the strongest fossil evidence supporting 
this scenario. This is beautifully described in Weinberg's book {\it The
First Three Minutes}\cite{weinberg1} published in 1977. 
It is remarkable that the work
over the last three decades has not found anything which would invalidate this 
view, but only strengthened the evidence for it. Furthermore, the 
subsequent research has made it possible to delineate the 
story beyond the first three minutes up
to the present, which was missing in Weinberg's book,
namely the story that is dominated by the formation of cosmic structure. 
Attempts to
understand cosmic structure formation have greatly enriched
cosmological tests both for structure formation itself and
for the evolution of the universe as a background to the structure.
Successful results of a number of key tests lead us to
conclude that we are approaching understanding of the 
evolution of the universe and cosmic structure.

Cosmology today is based on three paradigms: (i) the 
hot Big Bang and the subsequent
Friedmann-Lema$\hat{\rm \i}$tre expanding universe, (ii) the universe 
today being dominated by cold dark matter (CDM)\cite{cdm}, 
and (iii) the presence
of inflation in some early period\cite{guth}. (ii) is still hypothetical
and (iii) is even more so. Yet, we cannot construct a reasonable model
of the universe without the aid of
these two concepts. The most important implication
of inflation is the generation of density fluctuations over 
superhorizon scales, the presence of which is firmly established by
the observations of the cosmic microwave background (CMB) by 
the COBE satellite\cite{cobe}. 

The formation of structure basically reads as follows. At some early epoch
density fluctuations are generated adiabatically. The most promising
idea ascribes the origin to thermal fluctuations of the Hawking radiation
in the de Sitter phase of inflation, and these fluctuations are frozen
into classical fluctuations in the inflation era\cite{guthpi}.
The observed fluctuations are 
close to Gaussian 
noise with their spectrum usually represented as 
\begin{equation}
P(k)=\langle|\delta_k|^2\rangle\propto k^n\ ,
\label{eq:1}
\end{equation} 
where $n$ is close to unity.
This noise is amplified by self gravity in an expanding 
universe\cite{lifshitz}.
The fluctuations grow only when the universe is matter dominated and the
scale considered (i.e., $2\pi/k$) is within the horizon. Therefore, 
the perturbations are modified by a scale dependent factor as they grow 
\begin{equation}
P(k,z)=D(z) k^n T(k)\ ,
\label{eq:2}
\end{equation}
where $D(z)$ is the growth factor as a function of redshift, and
the transfer function $T(k)\sim 1$ for small $k$ and $\sim k^{-4}$ 
for large $k$.
The transition takes place at $k\simeq k_{eq}\simeq 2\pi/c t_{eq}$, 
where 

\begin{equation}
ct_{eq}=6.5(\Omega h)^{-1}h^{-1}{\rm Mpc}
\label{eq:3}
\end{equation}
is the horizon scale at matter - radiation equality\cite{bbks}. 
$T(k)$ is called the transfer function. Hence, the universe acts as
a low-pass gravitational amplifier of cosmic noise. 
The amplitude of the fluctuations
that enter the horizon is nearly constant ($n\simeq1$)\cite{HZ}
and is of the
order of $10^{-5}$. The small-scale fluctuations become non-linear
at $z\simeq 10-20$, and the first objects form from high peaks of
rare Gaussian fluctuations. As time passes, lower peaks
and larger scale fluctuations enter the non-linear regime,
and eventually form gravitationally bound systems which decouple from
the expansion of the universe. We call this state `collapsed'.
At the present epoch ($z=0$) objects larger than $\sim 8h^{-1}$Mpc 
are still in the linear regime. 

\begin{figure}
\epsfxsize160pt
\figurebox{}{}{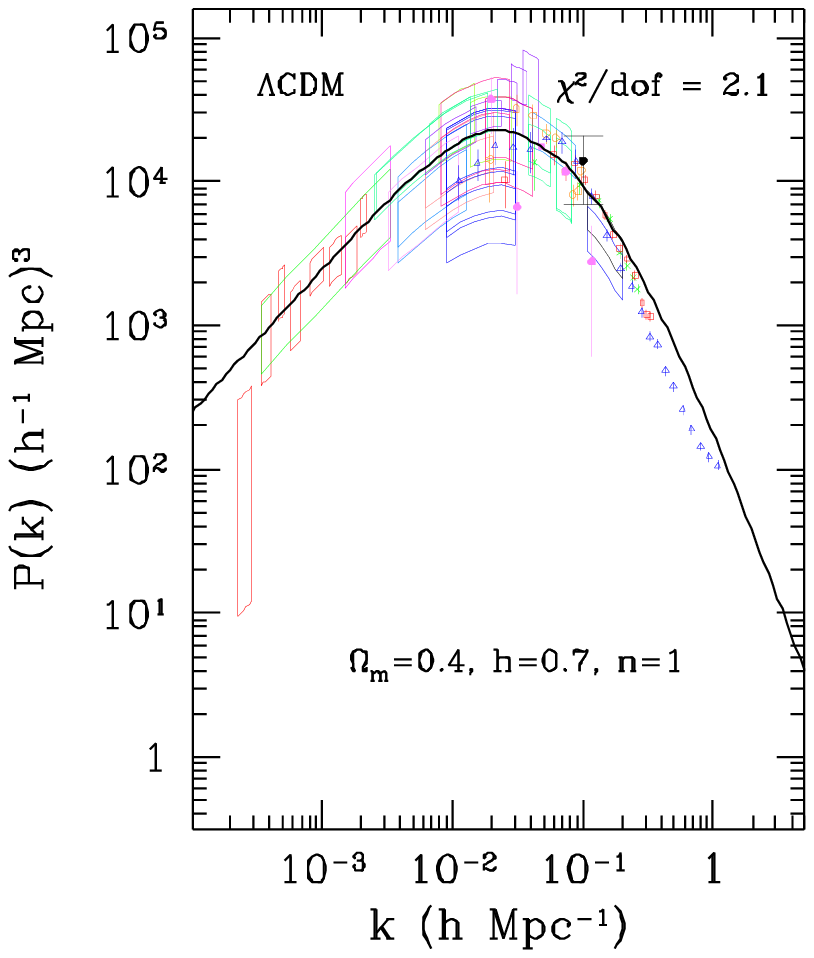}
\caption{
Power spectrum derived from large scale clustering of
galaxies (shown with data points) and CMB temperature fluctuations
(shown with boxes). The curve is the model power spectrum of a flat
CDM universe with a cosmological constant.
The figure is taken from Gawiser and Silk $^9$.
}
\label{figure1}
\end{figure}

The fluctuations and their evolution are described by a
single function of the power spectrum $P(k)$ scaled to today,
with the normalisation represented by rms mass fluctuations
within spheres of radius of 8$h^{-1}$Mpc:
\begin{eqnarray}
\sigma_8&=&\langle (\delta M/M)^2\rangle^{1/2}|_{R=8h^{-1}{\rm Mpc}}\\
        &=& \int^\infty_0 4\pi k^2dk |\delta_k|^2W(kR)\ ,
\end{eqnarray}
where $W(kR)$ is called the window function.
The fluctuations (adiabatic fluctuations) before recombination epoch
($z\sim 1000$) are imprinted on the CMB, and they are conveniently
represented by multipoles of the temperature field as,
\begin{equation}
\langle (\Delta T/T)^2\rangle=\sum_\ell{2\ell+1\over 4\pi}C_\ell\ .
\end{equation}
For illustration the power spectrum is shown in Figure~\ref{figure1}.

 The empirical
match of the power spectra estimated from large scale
galaxy clustering and
from CMB (COBE), which generically differ by a factor of $10^5$, 
has brought us confidence that we are working with
the correct theory\cite{wright,bew}. Here the CDM hypothesis
plays a crucial role. 
Without CDM this matching is impossible, or more precisely we do not know any 
alternatives yet.

\begin{figure}
\epsfxsize160pt
\figurebox{}{}{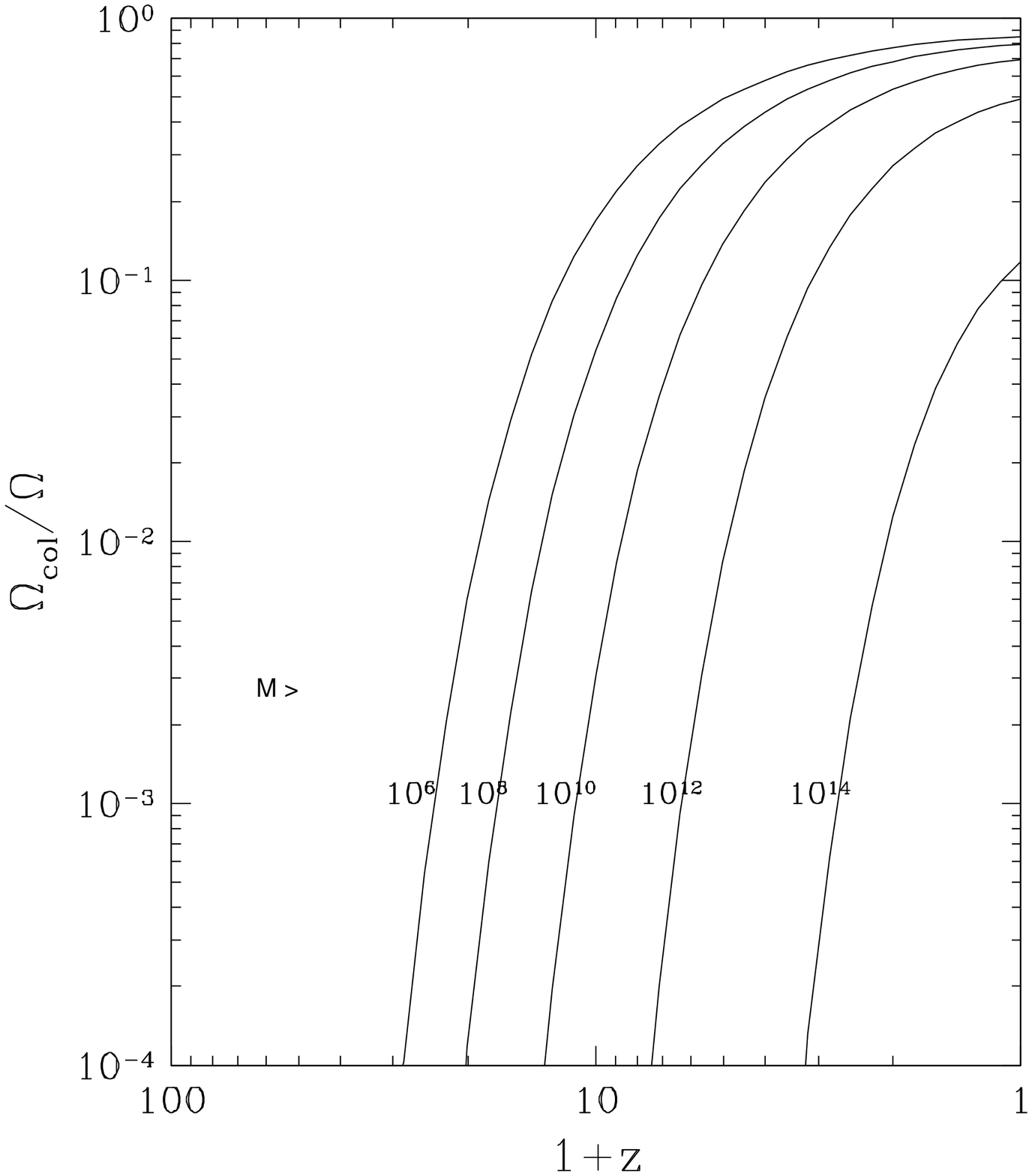}
\caption{
Fraction of gravitationally collapsed objects 
with mass greater than a given value $M$ (in solar mass units) 
as a function of redshift
$1+z$. The calculation uses the Press-Schechter formalism and assumes
the CDM model with parameters $\Omega=0.3$, $\lambda=0.7$
and $\sigma_8=0.9$.
}
\label{figure2}
\end{figure}

As fluctuations grow, they enter a non-linear regime.
This phase was first extensively studied by the use of $N$ body simulations.
The statistical results of simulations are very well described with 
an approach called the Press-Schechter formalism\cite{PS}, in which
Gaussian fluctuations, when they exceed some threshold\cite{kaiser,bbks}, 
follow 
nonlinear evolution as described by a spherical collapse
model\cite{gunngott}. This allows us to treat statistical aspects of
non-linear growth in an analytic way. Figure~\ref{figure2} 
shows the mass fraction
of collapsed objects with mass $>M$ as a function of redshift $z$.

Whether the collapsed object 
forms a brightly shining single entity (galaxy) or an
assembly of galaxies depends on the cooling time scale ($t_{\rm cool}$) 
compared to the dynamical scale, $t_{\rm dyn}\sim (G\rho)^{-1/2}$
\cite{cooling}.
For $t_{\rm cool}<t_{\rm dyn}$, the object cools and shrinks by 
dissipation and stars form, shining as a (proto)galaxy. Otherwise, 
the object remains a virialised
cloud, and is observed as a group or a cluster of galaxies. 
In the latter case
only gravity works efficiently, so that the system is sufficiently
simple to serve as a test for gravitational clustering theory.
Galaxy formation is very complicated due not only to the action
of the cooling process, which eventually leads to star formation, but 
also to feedback effects, such as UV radiation and supernova winds 
from stars. 
We expect that the first galaxies form at around $z\sim 10$. The period between
$z\sim 1000$ and the epoch of first galaxy formation constitutes a dark
age of cosmological history. Observationally, the highest redshift securely
measured is $z=5.8$ for a quasar\cite{z5.8}. Even higher redshift galaxies
have been reported, though the redshift measurement is not as secure 
as for quasars. How galaxies formed and evolved is the
most important arena for astrophysical cosmologists today, both
theoretically and observationally. I omit to discuss 
this subject in this
talk, however, since it does not seem to give us 
insights into particle physics; it
is entirely a world of astrophysics.

Before concluding this section I would emphasise that crucial 
cosmological tests can be made by the convergence of cosmological 
parameters, notably the Hubble constant, $H_0$, the cosmic matter density
$\rho$ in units of the Einstein-de Sitter (EdS) 
closure value $\rho_{\rm crit}$,  
$\Omega$, and the cosmological constant or vacuum energy density in units of
$\rho_{\rm crit}$, $\lambda=\Lambda/3H_0^2=\rho_V/\rho_{\rm crit}$; 
$\Omega+\lambda=1$ defines
a flat (zero curvature) universe. $H_0$ is often represented 
by $h=H_0/100$ km s$^{-1}$Mpc$^{-1}$.

\section{Hubble constant}

The Hubble  constant, which has dimension of inverse time,   sets the
scale of the size and age of the Universe.  Recent efforts have almost 
solved the long-standing discrepancy concerning the extragalactic 
distance scale; at the same time, however,  
significant uncertainties  are newly revealed in the distance scale
within the Milky Way and to the Large Magellanic Cloud (LMC), 
the mile stone to the extragalactic distance.

The global value of $H_0$  was   uncertain by a factor of
two for several decades. The discovery of a few new distance indicators
around 1990 has made possible an estimation of 
the systematic error for each
indicator by cross-correlating the resulting distances
(For a review of the methods, see [17]). This greatly enhanced
the reliability of Hubble constant determinations. The error shrunk,
notwithstanding there was still a dichotomous discrepancy between $H_0=80$
and 50 depending on the method used\cite{fhp1}. 
This meant that there still existed
systematic effects that were not understood.    
The next major advance  was brought with the refurbishment mission 
of HST, which enabled one to resolve
Cepheids in galaxies as distant as 20 Mpc (HST Key Project\cite{freedman1}). 
This secured the
distance to the Virgo cluster and tightened the 
calibrations of the extragalactic distance indicators, and
resulted in $H_0=(70-75)\pm10$, 10\% lower than the `high value'\cite{fmm}.
Another important contribution was the discovery that the 
maximum brightness of type Ia supernovae (SNeIa)  
is not an absolute standard candle,
but correlates with the decline rate
of brightness\cite{pskov-phil}, 
along with the direct calibration of the maximum brightness
of several SNeIa with HST Cepheid observations\cite{saha-sandage}.
The resulting $H_0$ was $64\pm3$, appreciably higher than 50.
These results nearly resolved the  long-standing controversy.

Extragalactic distance ladders are calibrated with the Cepheid 
period-luminosity relation. The majority of observations
are provided by the HST-KP, whereas
those of SNIa host galaxies are given by Saha, Sandage and collaborators. 
It was found by the HST-KP group that the Saha-Sandage distances that calibrate
the SNIa brightness 
are all systematically longer by 5-10\%\cite{gibson} for different reasons
for different galaxies\cite{freedman2}. The result of HST-KP is 
confirmed by [25].
This makes the Hubble constant from SNeIa $69\pm4$.  

Another distance indicator that allows an accurate estimation is 
surface brightness fluctuations. The current best result based on
300 galaxies is $H_0=77\pm7$, or $74\pm4$ with a model of the peculiar
velocity flow from galaxy density distributions\cite{blakeslee}.
Taking SN and SBF results we may conclude $H_0=71\pm7$ (2$\sigma$)\cite{fh}, 
in agreement with the 2000 summary of the HST-KP group\cite{mould}. 
HST-KP group slightly updated $H_0$ 
in their later report\cite{freedman3}: $H_0=74\pm7$.
Further reduction of the error needs accurate understanding of
interstellar extinction corrections and metallicity effects, which is 
by no means easy.

All the methods mentioned above use distance ladders and take the 
distance to the Large Magellanic Cloud  (LMC) to be 50 kpc 
(distance modulus $m-M=18.5$)
as the zero point. Before 1997 few doubts were cast on this. 
With the exception of RR Lyraes, the distances have converged to
$m-M=18.5\pm0.1$, i.e., within 5\% error, and the  
RR Lyr discrepancy was blamed 
on its larger calibration error.
The work over the last three years, notably that by the 
Hipparcos astrometric satellite (ESA 1997), revealed
that the distance to LMC is not as secure as has been
thought. The current estimate of the LMC distance varies 
from 43 to 53 kpc. This means that the Hubble
constant is uncertain by a factor 0.95$-$1.15 \cite{fh}.
Leaving this uncertainty I conclude 
the Hubble constant to be

\begin{equation}
H_0=71\pm7\times{1.15 \atop 0.95}~ {\rm km}~{\rm s}^{-1}{\rm Mpc}^{-1}\ .
\label{eq:4}
\end{equation}

\noindent
{\it Cosmic age}

The estimate of the age of the universe uses the position of turn off from
main sequence tracks in the HR diagram of globular clusters. 
The age thus estimated also
turned out to be more uncertain than had been thought. The major elements
of uncertainties are the zero point of RR Lyr (20\%) and
the interpretation as to the formation of globular
clusters, whether their heavy element abundances indicate the formation
epoch or their formation was coeval independent of the heavy element 
abundance (20\%). The first uncertainty is related to that of the LMC
distance: the calibration giving a long distance to LMC gives a shorter
age of clusters. The minimum of the estimated age is 12$\pm$1 Gyr\cite{age}
and the maximum is 18$\pm$2 Gyr, where $\pm$ reflects errors other than
are discussed here; see [27].

\section{$\Omega$ and $\Lambda$}

The mass density parameter $\Omega$, as measured in units of
the critical density, controls the cosmic structure formation. 
From the cosmic structure formation point of view the role of the cosmological
constant $\lambda$ is subdominant: it partially compensates the slow 
speed of structure formation in a low density universe.

Whether $\Omega$ and $\lambda$ inferred from the geometry of the universe
or dynamics agree with those with the aid of structure formation models
serves as an
important cosmological test not only for the validity of the
Friedmann universe but also
for the model of cosmic structure formation.

Determinations of  $\Omega$ and $\lambda$ which can be carried out without
resorting to specific structure formation models are:

\noindent
(1) $H_0-t_0$ matching using $t_0=H_0^{-1}f(\Omega,\lambda)$, which gives
$\Omega<0.8-0.9$. This at least excludes the Einstein-de Sitter 
($\Omega=1$) universe.

\noindent
(2) Luminosity density and the average mass to light ratio of galaxies, 
$\Omega={\cal L}\langle M/L\rangle/\rho_{\rm crit}$. This gives $\Omega=
0.2-0.5$. A slightly larger value compared to those in the literature 
is due to a correction for unclustered components\cite{f00b}.

\noindent
(3)$^*$ Cluster baryon fraction, as inferred from X-ray 
emissivity\cite{white1,WF} or
the Zeldovich-Sunyaev effect\cite{myers}. 
This should match with the global value $\Omega_b/\Omega$, where $\Omega_b$
is the baryon density inferred from primordial nucleosynthesis.

\noindent
(4) Peculiar velocity - overdensity relation, 
$\nabla\cdot v_p=-H_0\Omega^{0.6}\delta$,
a direct derivative from gravitational instability theory\cite{peebles2}. 
The result of this test is still grossly controversial; 
the estimate varies from
$\Omega=0.2$ to 1.  

\noindent
(5) Type Ia supernova Hubble diagram, which
measures the luminosity distance, $d_L=d_L(z;\Omega,\lambda)$. The result
is summarised as $\Omega=0.8\lambda-0.4\pm0.4$ \cite{high-z,sncp}.

\noindent
(6) Gravitational lensing frequency. The image of distant quasars 
occasionally splits into two or more images due to foreground
galaxy's gravitational
potential. The frequency is sensitive to $\lambda$, while it is nearly
independent of $\Omega$. The current limit\cite{f00b} is $\lambda<0.8$.

Determinations that depend on specific structure formation models are:

\noindent
(7) Shape parameter of the transfer function. The break of the transfer
function depends on the shape parameter $\Gamma\simeq\Omega h$, and this
is estimated from large scale galaxy clustering, as 
$\Gamma+(n-1)/2=0.15-0.3$ \cite{efs1,peacock}.
This means $\Omega\simeq 0.35$. 

(8)$^*$ Matching of the cluster abundance with the COBE normalisation. 
The cluster abundance at $z\approx 0$ requires the rms mass fluctuation 
to satisfy\cite{Ncluster} $\sigma_8\approx 0.5\Omega^{-0.5}$ .
$\sigma_8$ is also accurately determined by the large-scale 
fluctuation power imprinted in the CMB with the aid of eq.(2). 
The result depends on the power $n$, but notwithstanding
$\Omega>0.5$ cannot be reconciled with observations unless $n$ is largely
deviated from unity\cite{bew}. 

(9)$^*$ Multipoles of CMB temperature fields: the position of the 
acoustic peak is roughly proportional to
$\ell_1\approx 220[(1-\lambda)/\Omega]^{1/2}$.
A compilation of $C_\ell$ measurements favours 
$\Omega+\lambda\approx 1$ \cite{efs2,lineweaver}.

(10) Evolution of cluster abundance\cite{oukbir}. 
The evolution of rich cluster to $z\approx
0.5-0.8$ is more sensitive to $\sigma_8$, and one can separately
determine $\sigma_8$ and $\Omega$ by studying the evolution of the
cluster abundance. The results, however, are
at present dichotomous: $\Omega=0.2-0.45$ \cite{high-zcl1}
and $\approx 1$ \cite{high-zcl2}.

The items with asterisks will be revisited in the next section.
Among the ten tests, (5), (6) and (10) are particularly important tests for
the cosmological constant. We have omitted well-known
`classical tests' such as the number count of galaxies, angular-diameter
redshift relation, and the redshift distribution of galaxies, since these 
tests depends on the galaxy evolution, the understanding of
which is still far from complete.

The conclusion we can draw from this list is a gross convergence to 
$\Omega=0.2-0.5$ and indications for a finite $\lambda$ (SNIa
Hubble diagram and CMB multipoles). 
We shall see in the next section that these conclusions are 
corroborated by the new data of CMB observations, as seen in
Figure~\ref{figure6} below. 

\section{Impacts of the new CMB experiments}

The hot news of this year is the data release of two high resolution CMB
anisotropy experiments using balloon flights. One is BOOMERanG\cite{boom}
flight in Antarctica observing the southern sky and the other is
MAXIMA\cite{max}, a North American flight exploring the northern sky.
The two experiments have 
the beam sizes of the order of 10$'$, significantly smaller than that of COBE, 
and explore CMB multipoles for $\ell=26-625$ and $36-785$, respectively. 
These data cover both first and second peak regions, and
MAXIMA marginally reaches the third peak region.
The one sigma error is about 10$-$20\% for each data point,
and normalisation errors are 10\% (BOOMERanG) and 4\%
(MAXIMA). BOOMERanG gives more restrictive information for
low $\ell$.

BOOMERanG presents a map of CMB sky at 90, 150, 240 GHz (at which CMB
is supposed to dominate) and 400 GHz (at which dust emission 
dominates). 
The maps at the first three frequencies show patterns in a remarkable 
agreement, verifying that the fluctuations are indeed intrinsic to CMB.
On the other hand, the map at 400 GHz shows a completely different
pattern, correlated well with interstellar dust emission known from
infrared observations\cite{sfd}. The 400 GHz map 
agrees with a map obtained from the difference of the
two maps at 240 and 150 GHz. 

\begin{figure}
\epsfxsize160pt
\figurebox{}{}{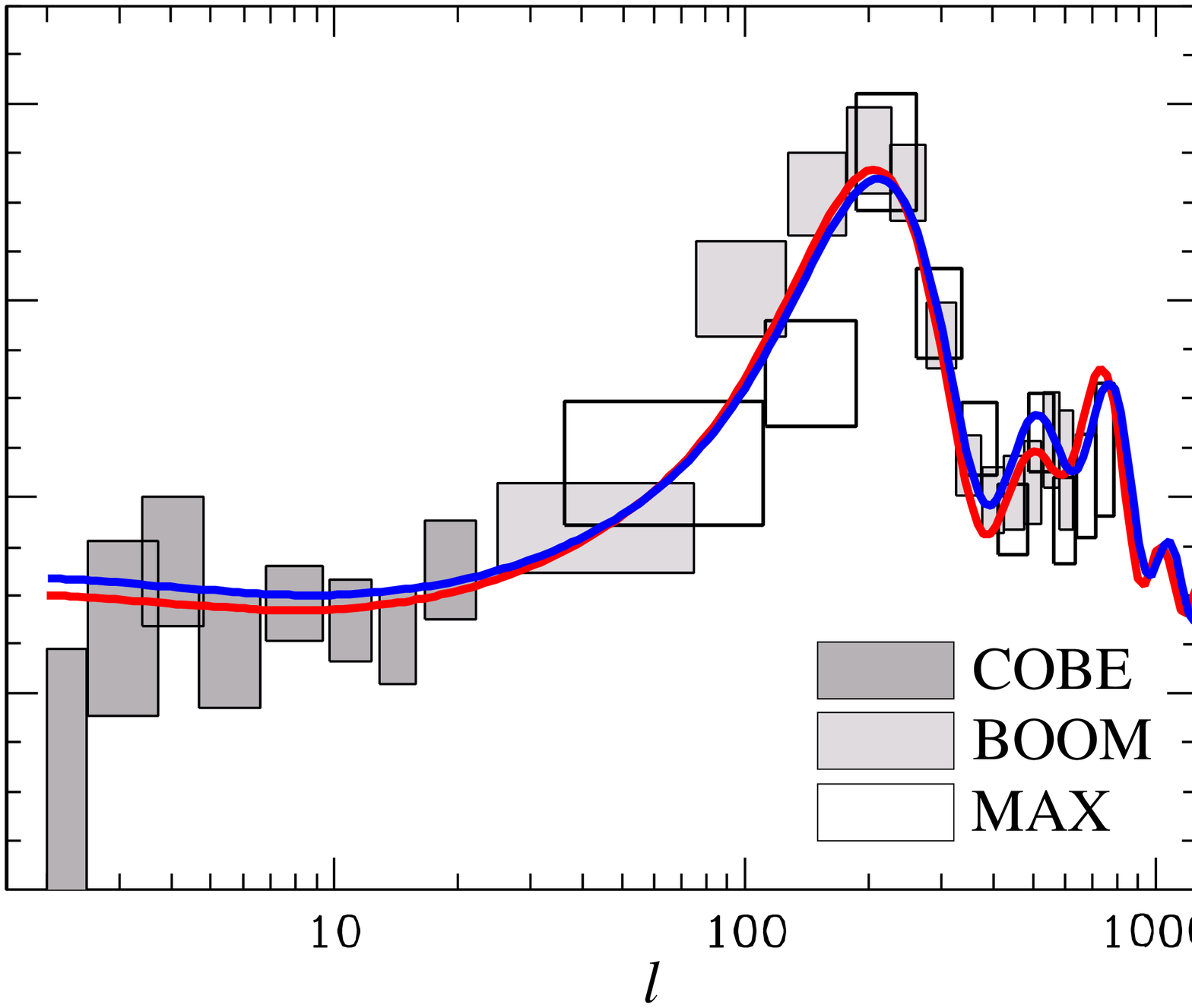}
\caption{
CMB multipoles 
$\Delta T_\ell=\sqrt{\ell(\ell+1)C_\ell/2\pi}$
from BOOMERanG and MAXIMA
experiments (the normalisations are 
shifted within one sigma calibration error), 
together with the COBE 4 year data. The curves show 
the prediction of the CDM structure formation model. The thick solid curve
represents the model that satisfies the joint constraint: $\Omega=0.35$,
$\lambda=0.65$, $h=0.75$, $\Omega_bh^2=0.023$ and $n=0.95$. The grey
curve is the model that is a good fit to CMB alone: $\Omega=0.3$,
$\lambda=0.7$, $h=0.9$, $\Omega_bh^2=0.03$ and $n=1$. Note a high
baryon abundance for the latter. Figure is taken from 
Hu et al.$^{53}$.
}
\label{figure3}
\end{figure}

The multipoles extracted from two observations, which observe 
the opposite sides of sky with respect
to the Galactic plane, show an excellent agreement 
with each other (except for one point
at $\ell\approx 140$) once we shift the data within 
the normalisation errors (see Figure~\ref{figure3}). 
These data are also quite consistent
with the earlier experiments if the average is taken over the data with
large errors and large scatters. The two experiments have
brought important improvement in the accuracy, which allows us to derive
solid conclusions on the cosmological parameters from CMB.

A number of extensive analyses followed the data release, and the
conclusions, while they are expressed in different ways, agree with
each other \cite{cmb1}$^-$\cite{cmb6}.
Most authors use a general likelihood analysis in multiparameter
(typically 7 parameter) space imposing varieties of prior conditions,
while Hu et al. \cite{cmb5} developed 
a parametric approach to make correlations among parameters more
visible.    

\begin{figure}
\epsfxsize160pt
\figurebox{}{}{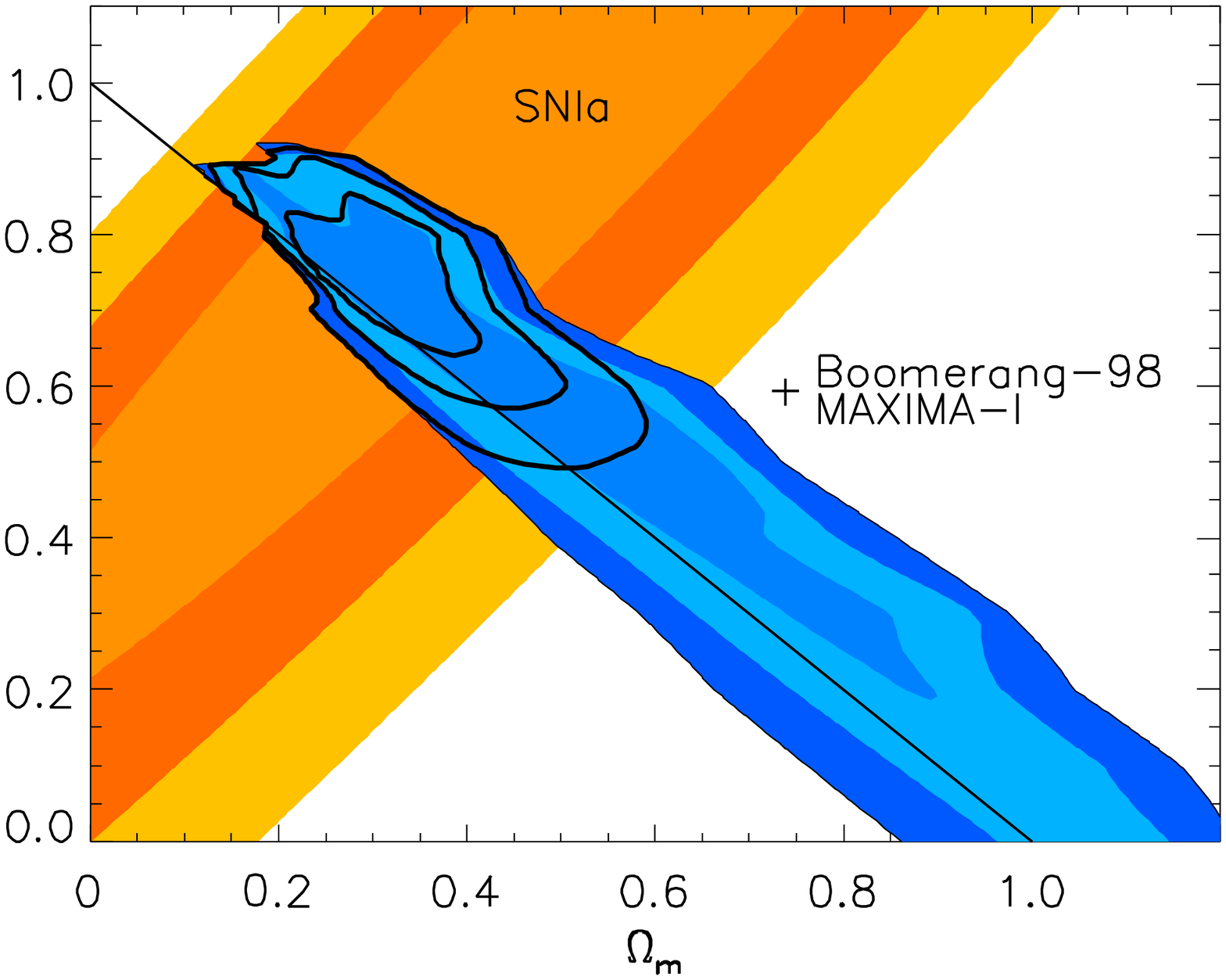}
\caption{
Constraints on the $\Omega-\lambda$ plane derived
from CMB multipoles and a type Ia supernova Hubble diagram, taken from
Jaffe et al.$^{54}$.
The labels are: $\Omega_m\equiv \Omega$ and 
$\Omega_\Lambda\equiv\lambda$.
The three levels of shading mean 1, 2 and 3 sigma. The contours
show the joint constraint. The straight line indicates flat universes.
}
\label{figure4}
\end{figure}

The major conclusions we can derive from these CMB data alone are:

\noindent
(1) The position of the first peak is securely measured to be $\ell=
206\pm 6$. This means that the universe is close to flat. 
See Figure~\ref{figure4}. 
See also (9) of section 3.

\noindent
(2) The spectral index is close to unity: $n=1\pm 0.08$.

\noindent
(3) The height of the second peak is significantly smaller than was expected
from the standard set of cosmological parameters, pointing to a 
baryon abundance that is higher than is inferred from primordial
nucleosynthesis\cite{hu}. The best fit requires $\Omega_bh^2\simeq
0.03$ compared with $\Omega_bh^2<0.023$ (95\% confidence) 
from nucleosynthesis.
The CMB data are consistent with 
the upper limit from nucleosynthesis only at a 2 sigma level
with a red tilt of the perturbation spectrum:

\begin{equation}
0.85<n<0.98.
\label{eq:8}
\end{equation}
The 2 $\sigma$ lower limit derived from 
CMB is $\Omega_bh^2>0.019$.

(1) is one of the most straightforward constraints derived from CMB,
and based only on geometry and acoustic physics. In the flat
universe the derived constraint agrees with $t_0<13.5$ Gyr.

(1) and (2) are taken to be a strong support for standard cosmology 
based on the CDM dominance of matter and adiabatic density perturbations.
They also support the idea of inflation as the origin of
density fluctuations; fluctuations from pure defects
(cosmic strings, textures) are excluded. 
On the other hand, (3) indicates
marginal consistency with the baryon abundance in our current standard 
understanding; inconsistency would become acute if the accuracy of the
data increases with the central values fixed.
 
\begin{figure}
\epsfxsize160pt
\figurebox{}{}{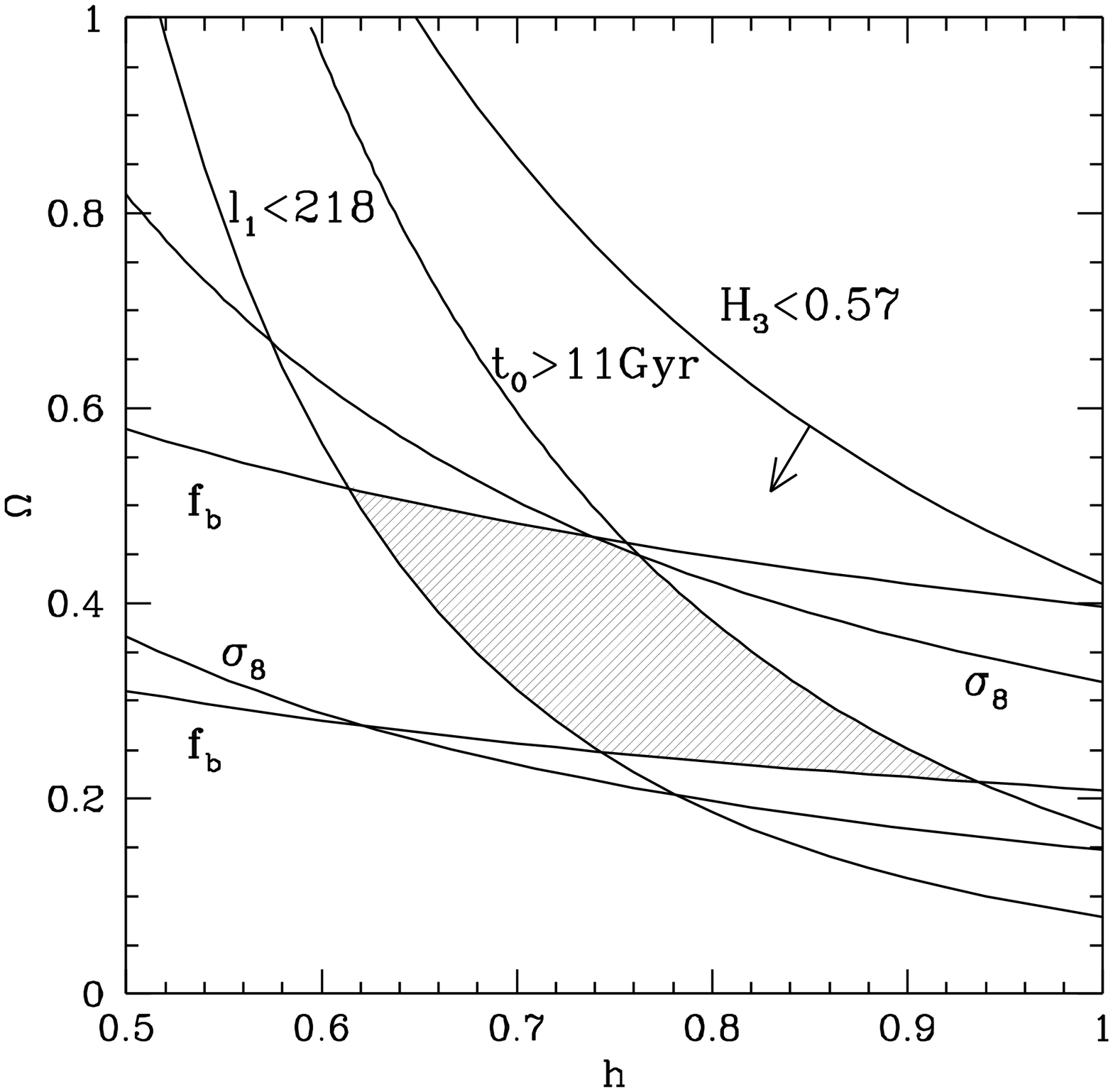}
\caption{
Constraints on the $h-\Omega$ plane derived
from the new CMB experiments. $\ell_1$ stands for the position of the
first peak, $H_3$ is the ratio of the heights of the third peak to the
second, the curves labelled by `$\sigma_8$' are obtained by the match 
of CMB data with the cluster abundance, and those with `$f_b$' are 
the constraint from the CMB and the cluster
baryon fraction. Cosmic age $t_0>11 Gyr$ is also plotted. The curves are
taken from [53].
}
\label{figure5}
\end{figure}

\begin{figure}
\epsfxsize160pt
\figurebox{}{}{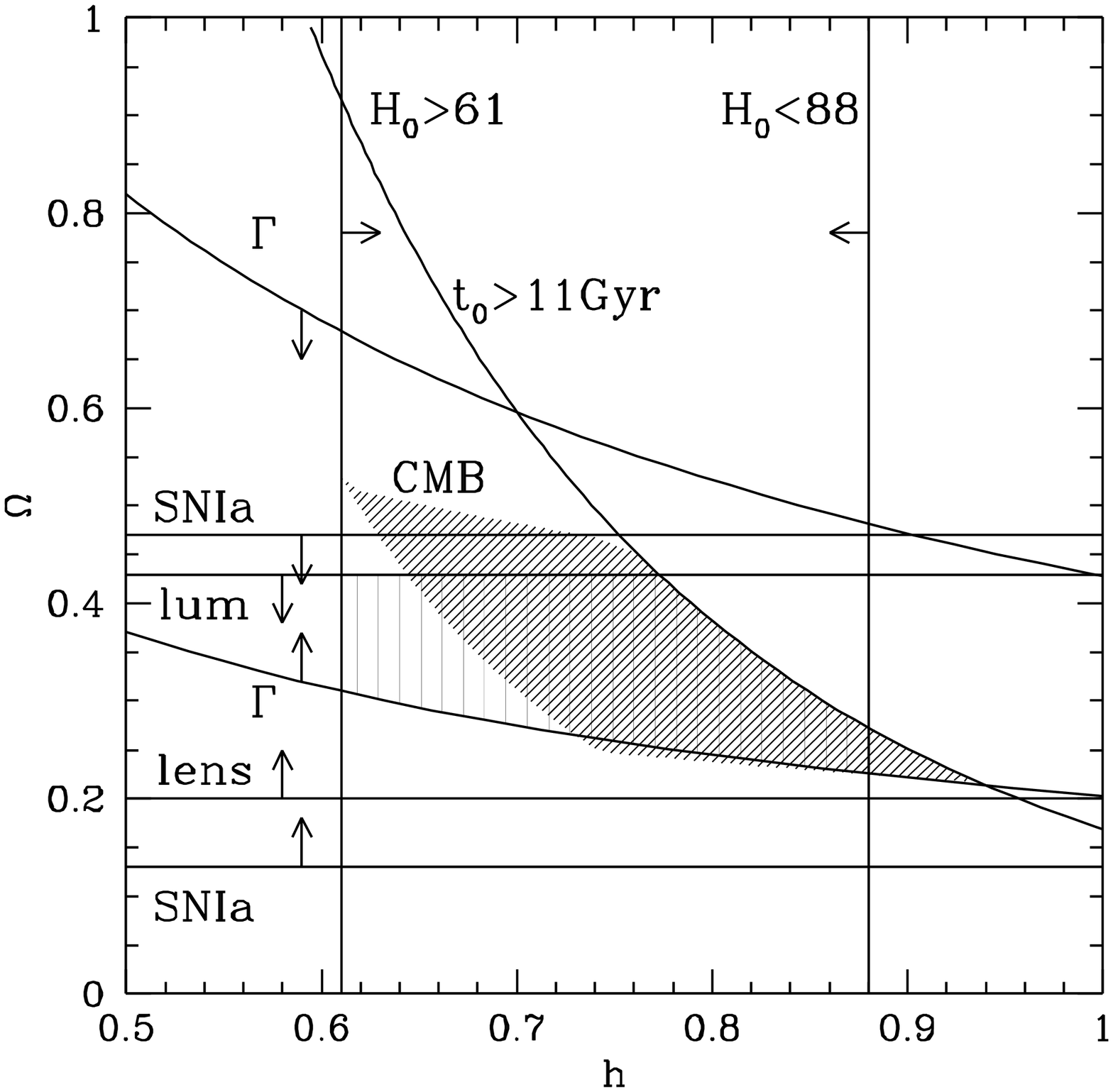}
\caption{
Summary of constraints shown in the $h-\Omega$ 
plane. Constraints shown by lines correspond to (1), (2), (5)-(7) 
of \S3 and the range of the Hubble constant in \S2. 
The allowed region derived from CMB (corresponding to (3), (8) and (9)
of \S3 and cosmic age) is shown by thick shading, while
those derived independent of CMB are indicated by light shading.
}
\label{figure6}
\end{figure}

The CMB data alone are not sufficient to uniquely determine the
cosmological parameters. When they are supplemented with the information
on large scale structure (either (7) or (8)), we are led to:

\begin{equation}
\Omega=0.4\pm0.2,~~~H_0=75\pm15,
~~~\lambda=1-\Omega{+0.2 \atop -0.1}\ .
\end{equation}
Figure~\ref{figure5} presents the constraints derived from the CMB either
directly or indirectly with the aid of external constraints 
($\Omega_b$ from nucleosynthesis, cluster abundance, and  cluster baryon
fraction) in the $h-\Omega$ plane.
The allowed parameter region agrees with what are discussed in 
section 3, as shown in Figure~\ref{figure6}, 
where it is shown together with
the constraints independent of CMB. The significance is that
the cosmological parameters derived via the structure formation model
agree with each other but also with model-independent analysis, 
corroborating our understanding of cosmology and structure formation.
This argument will be elaborated (or falsified) upon the completion of
the currently on-going CMB experiments, DASI, CBI and MAP.

\section{Matter content of the universe}

\subsection{Baryons}

The total baryon abundance, as represented by the baryon to photon ratio
$\eta=n_b/n_\gamma$, is inferred from nucleosynthesis
of light elements $d$, $^4$He, and $^7$Li. With $T=2.728$K, 
we have $\Omega_bh^2=0.00367(\eta/10^{-10})$.
A recent review of Olive et al.\cite{osw} gives two solutions 
$0.004<\Omega_bh^2<0.010$ and $0.015<\Omega_bh^2<0.023$ as 
2 sigma allowed ranges. The major change over the last five years is
the new input from deuterium abundance measured for Lyman $\alpha$ absorbing
clouds (Lyman limit systems) at high redshift, and a higher 
He abundance reported by Izotov \& Thuan\cite{izotov}. 
Deuterium lines are observed
for five Lyman limit systems; three of them gives low deuterium abundance,
while the other two (including the one observed at the first 
time\cite{songaila}) give high abundance. Assuming the lower abundance
to represent the true value,  
Tytler takes an average of the three and concludes
D/H$=3.4\pm0.25\times10^{-5}$, which turns into 
$0.019<\Omega_bh^2<0.021$ \cite{tytler}.
The primordial He abundance of Izotov \& Thuan is $Y_p=0.245\pm0.002$,
which is compared with the traditional value $0.235\pm0.003$. The
difference primarily arises from the use of different calculation of 
the helium recombination rates
and different corrections for collisional excitation than were used
in the past. So the difference is of systematic nature rather than due to
errors in the observations. It seems that more thorough studies of 
systematic errors are needed for the extraction of primordial
elemental abundances.

\begin{figure}
\epsfxsize160pt
\figurebox{}{}{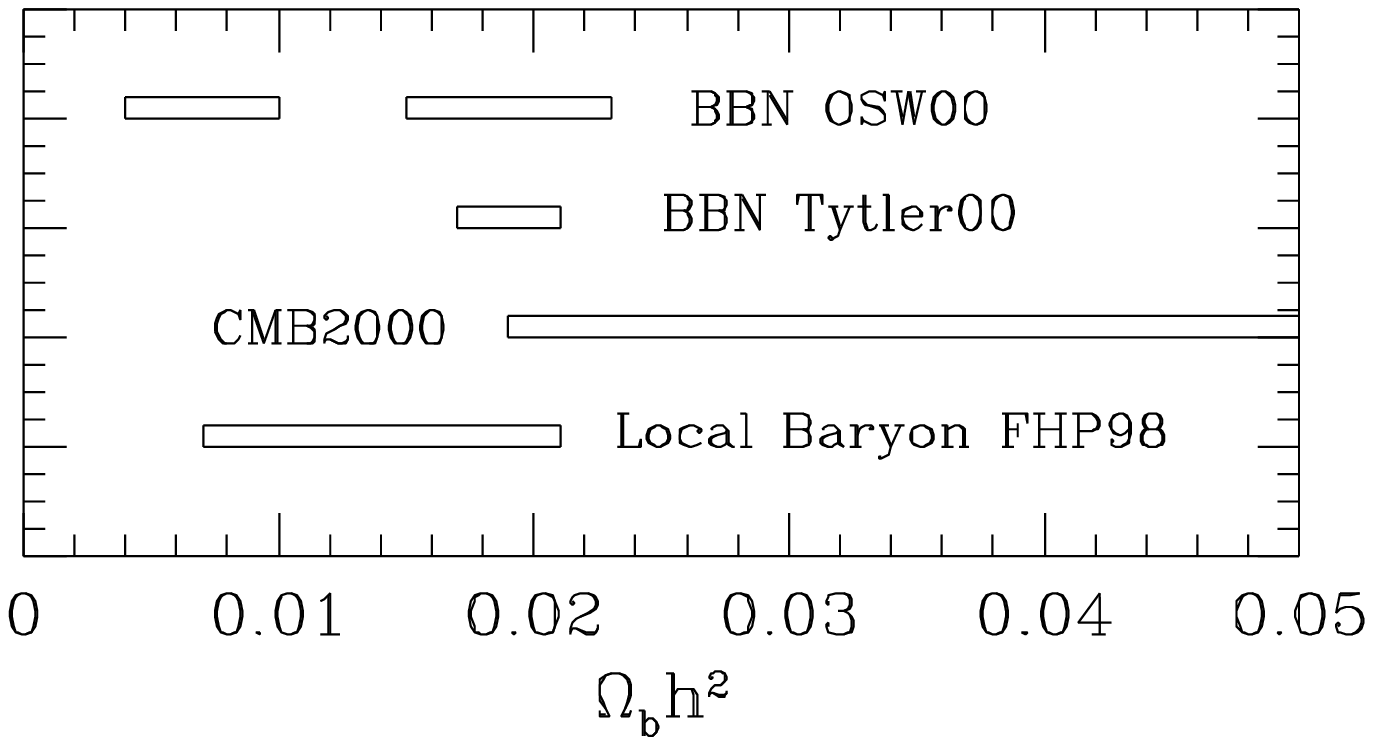}
\caption{
Baryon abundances inferred from CMB
(discussed above), 
nucleosynthesis (BBN)$^{56,59}$,
and accounting the local 
baryon distribution$^{60}$.
}
\label{figure7}
\end{figure}

The relative height of even to odd harmonic peaks of CMB multipoles
is sensitive to the baryon abundance, and the upper limit from BOOMERanG and
MAXIMA clearly rules out the low baryon option, but also it is 
marginally consistent with the high baryon option (see Figure~\ref{figure7}).

We note that only 10\% of baryons are frozen in stars which are visible in 
optical observations $\Omega_{\rm star}=0.004\pm0.002$ at $h=0.7$; 
baryons in the hot gas component which is visible through X-ray emission
is a similar amount. It is inferred that the rest is present around the   
galaxies in the form of warm gas that is not easily detectable. 
It seems that the high baryon option is barely
consistent with the amount which is obtained by adding 
all budget list for baryons\cite{fhp2}.

\subsection{Dark matter}

The presence of `non-baryonic' dark matter is compelling. Among others
the most important evidence is (i) the mismatch of $\Omega\approx 0.3$ 
with $\Omega_b$ 
from nucleosynthesis by one order of magnitude, and (ii) the matching of
fluctuations in CMB at $z=1000$ with those inferred from large scale 
structure at $z\approx 0$. If dark matter is baryonic
and couples to photons, this agreement of (ii) is completely lost: yet
we do not know any theories that give a correct matching between CMB and 
large scale structure without the aid of the CDM dominance.

A promising candidate of the dark matter is weak interacting massive
particles as a relic of the hot universe, as discussed widely by 
particle physicists (see talks by Drees, Olive and Arnowitt\cite{dm-osaka}). 
If these particles were in thermal equilibrium the dark matter density
would be
$\Omega\sim 3\times 10^{-27}
\langle\sigma_{\rm ann}v_{\rm rel}\rangle^{-1}$,
where average is taken over thermal distributions at the epoch of the
decoupling of dark matter, which is about $T\sim0.05m_{\rm dark}$.
The important fact is that the desired amount of dark matter is 
obtained with physics of typical weak 
interaction scale: $ \langle\sigma_{\rm ann}v_{\rm rel}\rangle^{-1}\sim
G_F^2T^2\sim 3\times 10^{-26}$ cm$^3$s$^{-1}$ for $m_{\rm dark}\sim 100$ GeV.
This makes the lowest supersymmetric particle a natural
candidate (see [62] for a review). 
The current most promising candidate is the 
neutralino that is a mixture of the photino, zino and Higgsino
(tg$\beta>3$, $M_\chi>50$ GeV)\cite{jellis,drees}.

\noindent
\underline{\it Massive neutrino}. 

\begin{figure}
\epsfxsize160pt
\figurebox{}{}{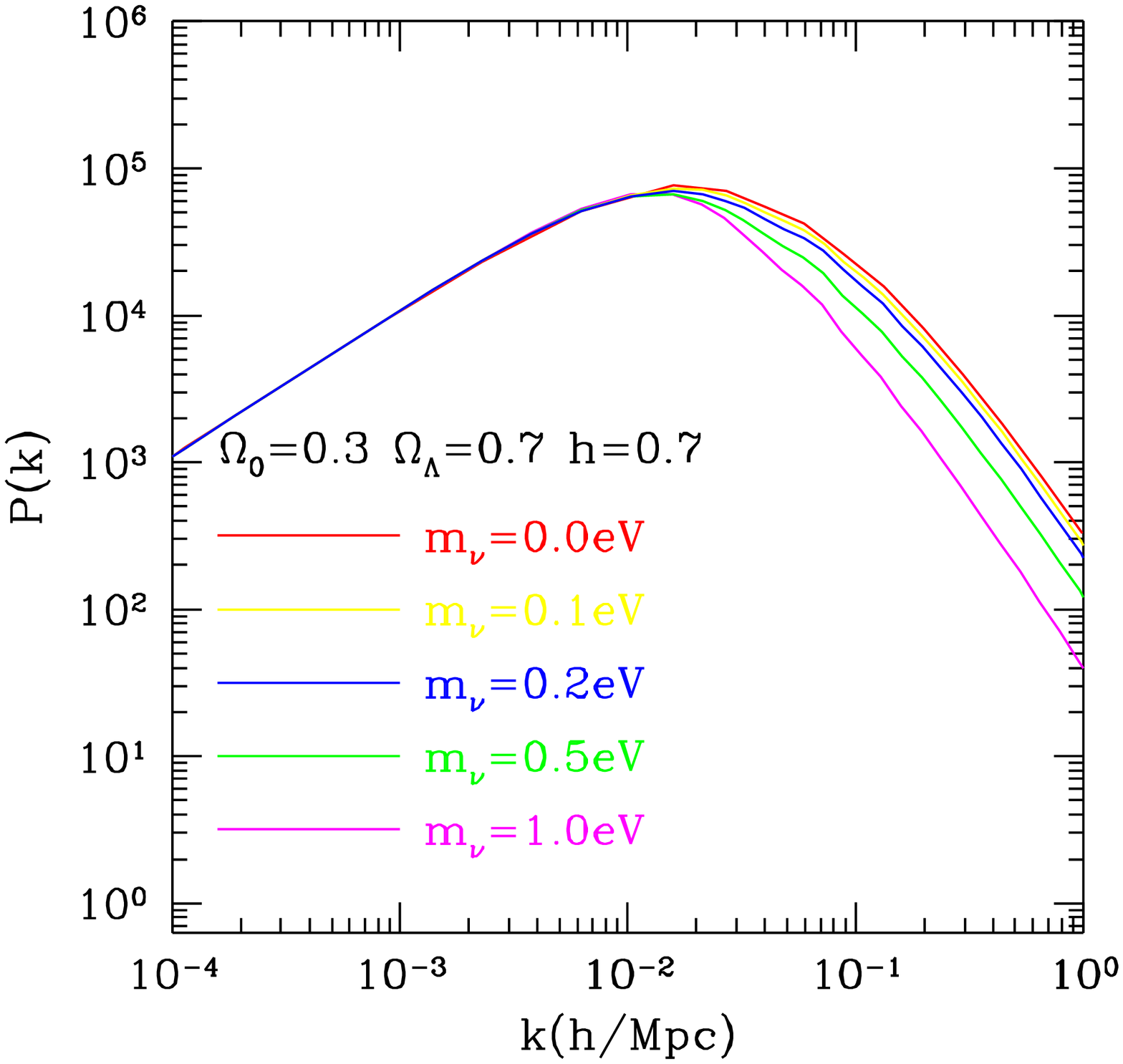}
\caption{
Effect of massive neutrinos on the power
spectrum. The three species of neutrinos are assumed to have equal mass, 
{\it i.e.},
$\sum_i m_{\nu_{i}}$ is three times the value indicated by labels.
The curve on the top is with $m_\nu=0$.
}
\label{figure8}
\end{figure}

We are now convinced that neutrinos
are massive. The mass density corresponding to $m_\nu\simeq0.05$ eV
is $\Omega_\nu\simeq 0.001$. This is the lower limit and 
the mass density could be larger if neutrino oscillation experiments 
are observing the difference of two or more degenerate neutrino masses.
From the view point of cosmology, neutrinos can no longer be 
a candidate for the dominant component of dark matter. The universe 
dominated by neutrinos does not give correct
structure formation, due to free streaming in the early universe
that smooths out small scale fluctuations. It has been discussed within
the EdS universe that a small admixture ($\sim 20$\%) of light neutrino
component would enhance relatively the large scale power 
required by observation\cite{mdm}. 
In a low matter density universe, however, sufficient 
large scale power is expected without massive neutrinos, and addition 
of massive neutrinos only disturbs the CMB cluster abundance 
matching\cite{het,vi,fls}.
Figure~\ref{figure8} shows the effect of massive neutrinos on the
power spectrum. The effect on small scale is apparent even if 
the neutrino mass is as small as 1 eV or less. Accepting the
conventional baryon abundance upper limit, the CMB-cluster abundance matching
leads to
\begin{equation}
\sum_i m_{\nu_i}<4 eV
\end{equation}
at 95 \% C.L.\cite{cmb5}.  A stronger limit is derived if mass density is
smaller, say $\Omega<0.4$ \cite{fls}.

\noindent
\underline{\it Strongly interacting dark matter?}
 
The possibility is recently discussed that dark matter might be strongly
interacting. The motivation is that $N$ body simulations with CDM 
models predict halo profiles more singular at the centre of the core
and more small-scale objects than are observed\cite{small-scale}.
This problem would be solved \footnote{Warm dark matter is another 
possibility\cite{warm}.} if 
dark matter is strongly interacting\cite{spergel} (see [71], 
however)
or it undergoes 
annihilation\cite{kkt}.
Either scenario requires the cross section to be of the order of
strong interaction. While this problem would offer another arena
for particle physics, it seems much more surprising if particles with such
properties exist in nature. I would like to ascribe it to our
incomplete understanding of astrophysics at small scales which is certainly
more complicated than large-scale dynamics.

\subsection{MACHO}

{\tt MACHO}'s (massive astrophysical compact halo objects) are a possible 
form of cold dark matter. These objects are collisionless, do not
couple to photons, and have no cutoff scale for perturbations; so
they would satisfy requirements for structure formation. 
The most likely candidate is aborted stars or stellar
remnants. This is nearly the only possibility for baryonic dark matter,
although the nucleosynthesis constraint on total baryonic matter still
have to be evaded. 

The novel feature of this
form of dark matter is that it would cause gravitational lensing when
it passes through the line of sight to background stars. The deflection
angle is too small to be observed, but it causes a large magnification
of the flux if it crosses the line of sight with a small impact 
parameter\cite{paczynski}.
The duration of `flash' is $\approx 70 (M/M_\odot)^{1/2}$ days for
background stars in LMC. 
Possible sources of confusion are variable or
flare stars that would mimic the effect. This confusion, however, can be
avoided by observing events with more than one wavelength passband,
since the stellar variation is associated with the variation
of temperature and the variations in different colour bands are not 
identical, in contrast to the gravitational lensing which is purely
geometrical.

Observations have been made mainly by two groups. 
The MACHO collaboration
conducted the observation for
5.7 years and found $13-17$ events towards LMC which are compared with
70 if the halo consists entirely of {\tt MACHO}\cite{macho}. 
The EROS group found 
3 events compared to 27 expected\cite{eros}. It is not clear whether detected 
`{\tt MACHO}'s are a part of the halo dark matter or not. From the 
time duration of the events
these objects should have mass between 0.1 to 1$M_\odot$, the 
mass typical of ordinary stars.
The MACHO collaboration
concludes the fraction $f$ of {\tt MACHO} to the total halo matter 
to be about 0.2,
while EROS group only quotes $f<0.4$ (at 95\% C.L.) as an upper limit.
The definitive and important conclusions 
are that (a) objects with a mass range of $10^{-7}-10^{-2}M_\odot$ cannot be
a candidate of dark matter, and (b) {\tt MACHO}, if any, accounts only for 
a minor fraction of dark matter. We add a remark that the lensing 
cannot be associated with main sequence stars. The number is also too large
if they are to be ascribed to white dwarfs 
(see [77] in this connection).

\section{Theory of cosmological constant}

We have now almost compelling evidence for a non-vanishing
cosmological constant. The traditional problem is why the cosmological
constant is so small, but now we are faced with anther problem why
it is non-vanishing and it is close to matter density. 
Traditional attempts to understand
the first problem are summarised by Weinberg\cite{weinberg2}. 
Many further attempts have recently been made, but the situation 
seems still far from the solution. Here I quote three categories of 
the attempts.

\noindent
(1) Time varying cosmological constant (Quintessence).
 
The cosmological constant is decreasing with time, as realised, for
example, by a scalar field (quintessence field)
slowly evolving down a potential. In this
view, we can start with a large cosmological constant.
The problem is why vacuum energy density and matter density are approximately
equal. Peebles \& Ratra\cite{PR} have considered the potential of the
form,

\begin{equation}
V\sim M^{4+\alpha}\phi^{-\alpha}\ .
\end{equation}
This model was studied more recently by Zlatev, Wang \& Steinhardt\cite{zws}
who showed that the $\phi$ field rolls down this potential works as
an attractor-like solution to the equation of motion, in the sense that
the field and its derivative approach a common evolutionary track
for a wide range of initial conditions (`tracker solution'). With an
appropriate choice of $M$ we get $\rho_V\sim\rho_m$. The weak points
are that one has to tune $M$ to give
$\rho_V\sim\rho_m$, and an addition of a constant term spoils the
desired behaviour. Armendariz-Picon et al.\cite{ams} further developed
a model so that negative pressure automatically becomes effective after
the epoch of matter-radiation equality. This model needs a modification
of the kinetic term ($k$-essence).

\noindent 
(2) Use of exact symmetry.

With exact supersymmetry the cosmological constant vanishes.
In supersymmetry theories in 4 space-time dimension, however, the disparity of 
fermion masses and their boson partners means breaking of supersymmetry,
which inevitably results in a positive cosmological constant of the order
of the supersymmetry breaking scale. This constant may be cancelled by 
some counter term in supergravity theory with an extreme fine-tuning, but 
this does not solve the problems posed above.
 
Witten observed in 3 dimensional theory that the disparity arises between
fermion and boson masses when matter interacts with gravity, while
supersymmetry is maintained\cite{witten}. Namely, $Q_{\rm SUSY}|0\rangle=0$
and $[Q_{\rm SUSY},H]=0$, so that the cosmological constant vanishes. When this
model is embedded into supergravity theory having a 
dilaton field, the compactified dimension is stretched to the
fourth dimension in the strong coupling limit of the dilaton coupling,
making the theory full four dimensional.

Once we have zero cosmological constant we must consider a mechanism to
generate a small vacuum energy. An idea is to consider 
ultra mini-chaotic inflation. If the potential for a scalar field
is very flat with the mass
of the order of Hubble constant, the initial state at $\phi\approx M_{\rm pl}$
still remains at the same value due to the Hubble viscosity, giving a
very small vacuum energy. Such a minuscule mass may be generated by
electroweak instanton effects. This proposal in [83] is justified by an
explicit calculation within a supergravity model\cite{watari}, which yields
$m\sim G_F^{5/4}m_q^{5/2}M_{\rm SUSY}^{3/2}M_{pl}^{-1/2}\exp(-4\pi^2/g^2_{2})$
($G_F$ is the Fermi constant, $m_q$ the quark mass, $M_{\rm SUSY}$ is
the SUSY breaking scale, and $g_2$ is the
SU(2) gauge coupling constant).
This gives about the correct
order of magnitude for the cosmological constant.

\noindent
(3) Anthropic principle.

If our universe is one member of an ensemble, and if the vacuum density
varies among the different members of the ensemble, the value observed
by us is conditioned by the necessity that the observed value should be
suitable for the evolution of our Galaxy and of intelligent life. This 
argument is called the anthropic principle, as explicitly stated by
Carter\cite{carter}. 
(More discussion will be made in section 8.)
An application of the anthropic principle to the vacuum energy is 
first discussed by Weinberg\cite{weinberg3}.

The argument is that galaxies would not have formed if vacuum energy were
larger than some critical value, since vacuum energy, providing 
a repulsive force, hinders evolved perturbations from collapsing into
galaxies. The condition is roughly expressed as $\rho_V<\rho_m$ at
$1+z\sim 4-5$, where galaxies formed. This translates to  
$\rho_V<100\rho_m$ today, a dramatic narrowing of the allowed range for
the cosmological constant.

Some authors further elaborated the argument by adopting the hypothesis called
`principle of mediocrity' which says that we should expect to find
ourselves in a big bang that is typical of those in which intelligent
life is possible\cite{vilenkin}. A working assumption to calculate
the probability of civilisation is that it is proportional to the 
number of baryons frozen into a galaxy, and the {\it a priori} probability
of a universe having $\rho_V$, $P(\rho_V)\simeq$constant.
Calculations are made for
galaxy formation according to the formalism described in section 1.
The result depends on further input assumptions, but it generally 
is not too far from the value with which we actually 
live\cite{cc-anthr}. For instance Martel et al. obtained the probability
of finding ourselves in a universe with $\lambda<0.7$ to be 5-12\%. 
Whether  $P(\rho_V)\simeq$constant is realised
is investigated within scalar field theory
in [89].

\section{Theory of inflation}

Cosmological inflation gives a universe a number of desirable features.
The most important among them is the generation of density fluctuations
over superhorizon scales. Quantum particle creations in the de Sitter
phase generate thermal fluctuations corresponding to the effective 
temperature $T=\hbar H/2\pi$ ($H$ being the expansion rate)
and they are frozen into classical fluctuations
when the scale considered goes outside the horizon in the inflation era.
Inflation is the only known mechanism that can
generate empirically viable fluctuations.

Theory of inflation assumes the presence of one or more scalar fields, 
called inflatons, which obey the field equation,

\begin{equation}
\ddot\phi+3H\dot\phi+{\partial V\over \partial \phi}=0\ .
\end{equation}
%
Inflation takes place if the second term that works as viscosity 
is large enough so that
roll down of the state is sufficiently slow: 
This slow roll regime is realised when 
\begin{equation}
\epsilon={M_{pl}^2\over16\pi}(V'/ V)^2 \ll 1,~~~
  \eta={M_{pl}^2\over8\pi^2}(V''/V) \ll 1 .
\end{equation}

There are several points to 
be satisfied to make the model observationally viable:

\noindent
(1) $\Omega+\lambda=1$

\noindent
(2) $N_{e{\rm-fold}}\ge 50$, where $N$ is the logarithmic ratio
of the scale factors before and after inflation. 

\noindent
(3) $V^{3/2}/M_{pl}^3V'=4\times 10^{-6}$ 
as required by the COBE observation, $Q=2\times 10^{-5}$. 
This is
a crucial condition to have successful structure formation and
the presence of ourselves.

\noindent
(4) The spectral index $n\approx 1$. The CMB observations indicate
$n=1\pm 0.15$ at 95\% C.L. If we require the consistency with 
primordial nucleosynthesis, only red tilt is allowed as shown 
in eq.(\ref{eq:8}).
%
The theory of inflation predicts
$n=1-6\epsilon+2\eta$ which is close to, but  
{\it not} quite unity.
Simple (one field) models of inflation predict 
$n<1$ (red tilt), whereas $n>1$ requires
a more complicated class of models, such as hybrid inflation. 

\noindent
(5) Tensor modes. Inflation may generate tensor perturbations, which
contribute to the CMB fluctuation for small $\ell$. 
Excessive tensor modes cause
the CMB harmonics increasing too rapidly towards a small $\ell$. 
A reasonable guess for the limit from the current data is $r=T/S<0.5$, but
detail statistical analyses are yet to be carried out. Inflation
that takes place significantly below the gravity scale does not
generate tensor perturbations\cite{lyth}. 

It would be instructive to impose our constraints on models
of inflation. For example, chaotic inflation with the potential
$V=m^2\phi^2$ (mass term only) predicts $n=0.96$ just consistent with
the upper limit of eq.(\ref{eq:8}). 
The value of the $\phi$ field at the epoch that the physical
scale goes out of the horizon is $\phi_{\rm phys}=2.8M_{pl}$ 
from (2). Inflaton mass $m=2\times 10^{13}$ GeV from (3). The model predicts
$T/S=0.12$, which is consistent with the observation.

There is a generic prediction of slow-roll inflation, $T/S\approx-6n_T$
with $n_T$ the spectral index of the tensor mode. Unfortunately, this
$n_T$ is the quantity most difficult to measure. The relation
\begin{equation}
T/S \approx 6(1-n)
\end{equation}
often quoted in the literature holds only for specific classes of
models. It is argued that most models of inflation predicts
the relation either close to eq.(14) or $T/S\approx 0$ \cite{HT}.

Many hundreds of inflation models have been considered by now\cite{inflation}.
I do not intend here to discuss model building, but it seems that
there are no satisfactory models.
I only briefly mention
the outline of models which seem more generic and why they are not
satisfactory; see [92] for details.

(1) $\phi^\alpha$ potential. This is the prototype for chaotic 
inflation\cite{linde1}. 
The slow roll condition and $e$-fold require that $\phi_{\rm phys}$ 
be larger than a few times $M_{\rm pl}$ and $\alpha$ be reasonably small. 
A red tilt $n<1$, and appreciable tensor perturbations that nearly
satisfy eq.(14) are predicted. 
We must deal with super-Planck scale physics, which
is beyond the understanding of particle physics today. The
real problem, however, is that there is no principle to forbid higher
order terms of the form $\phi^n/M_{\rm pl}^{n-4}$, which spoils
the slow roll condition for $\phi\gsim M_{pl}$. 
An attempt to forbid such terms by
introducing symmetry is presented
at this Conference\cite{kawasaki}, but it lacks 
a particle physics motivation.

(2) $V_0[1-({\phi\over\mu})^p]$ type potential. This is typical of
 `new inflation'\cite{ASL}. Inflation starts with $\phi\simeq 0$,
and ends with $\phi_{\rm end}\ll M_{pl}$.
A red tilt is predicted, while tensor perturbations are very small due
to a low energy scale involved.  
The difficulty is that one needs fine tuning for the initial condition.
Furthermore, in most models of this type the inflaton field is not 
in thermal equilibrium, so that symmetry restoration at high 
temperature does not work.
There is also a fundamental problem as to why universe has not
collapsed long before the onset of this inflation. 

(3) Hybrid inflation with two fields. The model is
$V(\phi,\sigma)=\lambda(\sigma^2-M^2)^2+m^2\phi^2/2+g\phi^2\sigma^2/2$. 
This is a model which combines
chaotic and new inflation features\cite{linde2}. For $\phi$ greater
than some critical value $\phi_c$, $\sigma=0$ is the minimum and
the model behaves as the chaotic type; $\phi$ remains large for a long
time. At the moment when $\phi$ becomes smaller than $\phi_c$, 
symmetry breaking occurs and rapid rolling of the field $\sigma$ 
takes place. One nice feature with this model is that it can 
be embedded into SUSY or supergravity models, and 
the energy scale
of the phase transition appears to agree with the unification 
scale\cite{copeland}.  
The problems have been pointed out in more recent studies, however, that
the model needs spatial homogeneity in the superhorizon scale
in the preinflation era\cite{vachaspati} and that the choice 
of the initial condition
needs fine tuning to keep the $\phi$ filed in the desired 
valley\cite{tetradis}. 
This model predicts blue tilt $n>1$ in the tree level, but the tilt
can be blue or red after loop corrections,
depending on input parameters\cite{inflation}. 
 
The general problem with inflation is a lack of satisfactory
models motivated from particle physics. For example, such an   
idea that simply combines inflation with supergravity theory is liable to fail 
because the K\"ahler potential is too curved with the exponential 
dependence of the field.
Most models discussed in the literature are constructed without 
regarding low energy physics;
so the models are those just to do it for its own sake alone.
For the view pint of astrophysical applications, the 
discovery that inflation does not exclude
open universes\cite{linde4} greatly diminished its predictive
power.
The observation of the tilt and the strength of the 
tensor mode will offer an important test for the model, though
the current data are not yet accurate enough for this purpose. 
Astrophysicists may not feel comfortable, however, unless particle physics
would explain why $V^{3/2}/V'$ takes a specific value as referred to in
(3).
A misprediction by an order of magnitude leads to a 
disaster for us (see below).

One philosophically interesting consequence arises from
the fact that inflation never ends (eternal inflation)
whichever inflation one considers\cite{linde5}. 
This would result in different
patches of the universe expanding differently; inflation leads to
great inhomogeneity at superhorizon scales. Many universes are born at
arbitrary instants in many different patches; after all we are living in
just one of them and observe this `small' patch as an `entire universe'.
This `multiverse' picture would give a base to the speculation that 
many physical 
parameters may vary in 
different universes. In this picture the Big Bang is no longer given
any special position.

\section{Anthropic principle: use or misuse?}

There are many constants that appear to be so tuned that they are
just appropriate for the evolution of intelligent life. We would
wonder whether what we expect to observe is restricted by the 
condition necessary for our presence as observers\cite{carter}.
We have discussed that the vacuum energy is one such example. This is also
true with the matter density. If the lightest neutrino would have 
mass larger than
10 keV, $\Omega_\nu\sim 100$ and the age of the universe would be too
short for
intelligent life to have developed. If, on the other hand, $\Omega<0.01$,
say, the galaxies would not have formed. In fact, 
$\Omega_{\rm  CDM}\sim\Omega_b\sim\Omega_\nu\sim\lambda\sim O(1)$ 
up to only three orders of magnitude is an intriguing coincidence. 

Another cosmologically important parameter is the strength of initial density
fluctuations $Q\sim O(10^{-5})$. If this were larger by one order of 
magnitude, galaxies would be dominated by vast black holes; no stars
or solar system could survive.
If it were smaller by one order of magnitude, cooling does not efficiently
work, and galaxies would not have formed\cite{rees}. From a view point of
particle theory this is an obscure quantity $\sim V^{3/2}/V'$ 
in terms of the inflaton potential.
Why this quantity takes a specific value which makes {\it us} habitable
is puzzling.

There is a similar tricky coincidence (or providence) also in
particle physics parameters, which is crucial
to the evolution of intelligent life. The central issue is the parameters
that affect element synthesis in the early universe and in stars.
A small change of quark mass and/or the QCD strength stabilises or
destabilises neutron, proton, deuterium, di-proton or di-neutron.
Furthermore, the production of elements heavier than carbon just
depends on the luck of the existence of a resonance in the
$^{12}$C system, which makes the bottleneck nuclear reaction 
$3\alpha\rightarrow^{12}$C possible. A similar situation also exists
with  $^{16}$O. Agrawal et al.\cite{agrawal} focused on the aspect that
weak interaction scale is close to QCD scale rather than the Planck
scale.  Hogan\cite{hogan} argued for the arrangement of mass difference
among $m_u$, $m_d$ and $m_e$, and a $\pm$1 MeV change of $m_d-m_u$ 
would disturb the existence of complex elements. He radically claims that
the correct unification scheme should {\it not} allow calculation of
$(m_d-m_u)/m_{\rm proton}$ from first principles. Rees\cite{rees} formulated
the requirement in a way that fractional binding energy of helium,   
$\epsilon={\rm BE}(^4{\rm He})/M(^4{\rm He})$, be tuned 
between 0.006 and 0.008. 
 
It is clear that our existence hinges on delicate tuning of many
parameters irrespective of whether it is a result of 
the anthropic principle or not. I refer the reader to Rees' book\cite{rees} 
for more arguments. Of course, 
the view on the anthropic principle is wildly divided.
Hawking considers that why we are living in 3+1 dimension but not in 2+1 or 2+2
and why low energy theory is SU(3)$\times$SU(2)$\times$U(1) etc. are all
results of the anthropic principle\cite{hawking}. 
Physicists usually hope that all parameters are derived up to 
only one from
fundamental principles, and the anthropic argument appears
for them to be equivalent to giving up this effort. 
For instance, Witten\cite{witten} states that ``I want to ultimately
understand that, with all the particle physics one day worked out, life is
possible in the universe because $\pi$ is between 3.14159 and 3.1416.
To me, understanding this would be the real anthropic principle ...''
It is disappointing if the anthropic principle is the solution to
many problems, but such a possibility is not excluded.

\section{Baryogenesis}

Before concluding this talk let me mention briefly baryogenesis,
which is in principle in the interface between cosmology and particle 
physics. The real contact between the two disciplines, however, 
is a subject in the future: astrophysical cosmologists argue about
an error of 10\% for the baryon abundance, whereas
particle physicists struggle to understand the order of magnitude.

Four major scenarios proposed so far and their state-of-the-art are:

(1) Grand unification.  This prototype baryogenesis idea does not receive 
much support.
Unless baryogenesis takes place with $B-L$ violated, the baryon excess is
erased above the electroweak scale under the effect of  
Kuzmin-Rubakov-Shaposhnikov's (KRS) sphalerons. 
With the presence of inflation,  
whether the reheating temperature is sufficiently high to produce
coloured Higgs is also a non-trivial problem. 
In SUSY GUT the reheat temperature $T_R$
needs to be $>m_H^c\sim 10^{17}$ GeV, which is the lower limit
on $m_H^c$ to avoid fast proton decay. In supergravity 
theory the reheat temperature cannot be sufficiently high 
($T_R<10^9$ GeV) to avoid copious gravitino production.

(2) Electroweak baryogenesis with the KRS effect. The necessary 
condition is that the electroweak phase transition is
of first order. Within the standard model this requires the Higgs mass
to be lower than 70GeV, which is already much lower than 
the current experimental limit. The possibility is not yet excluded
in supersymmetric extension. The electroweak transition can be 
strong if the stop mass is lower than the top quark mass\cite{carena}.
A possible large relative phase between the vacuum expectation values
of two Higgs doublets may bring CP violation large enough
to give the observed magnitude of baryon number. 
 
(3) Leptogenesis from heavy Majorana neutrino decay and 
the KRS mechanism. This mechanism works in varieties of unified models
with massive Majorana neutrinos. For a recent review see [107]. 
Another mechanism is proposed for leptogenesis via neutrino 
oscillation\cite{rubakov}.

(4) Affleck-Dine baryo/leptogenesis\cite{DA}. When the flat direction
of the SUSY potential is lifted by higher-dimensional effective
operator, coherent production of slepton and squark fields that carry baryon 
and lepton number takes place in the reheat phase of inflation.
The oscillation starts and ends earlier than was thought in the
original paper due to thermal plasma effect\cite{dine,allah,yanagida},
leading to some suppressions of the baryon or lepton number production.
Notwithstanding, this is still a viable scenario, although proper
treatment of 
leptogenesis requires the lightest neutrino mass to be smaller than
$10^{-8}$ eV for $T_R<10^9$ GeV \cite{yanagida}, the limit being
stronger than was obtained in [110]. 

\section{Conclusion}

Over the last few years our understanding of cosmic structure formation
based on the CDM dominance and statistical description has significantly 
tightened.  
The new CMB experiments reported this year further corroborated
the validity of the model. 
Concerning the world model of the universe we may conclude that
(1) open universes are excluded, (2) the EdS universe is excluded,
and (3) a non-zero cosmological constant is present. These conclusions
seem to be compelling in so far as we keep the current cosmic
structure formation model. Note that the CDM model is the only model known
today that successfully describes widely different observations.
The cosmological
parameters are converging to $H_0=62-83$ km s$^{-1}$Mpc$^{-1}$, $\Omega=
0.25-0.48$ and $\lambda=0.75-0.52$.

For astrophysical cosmology an interesting problem is the baryon 
abundance. The CMB experiments indicate the optimum value of the
baryon abundance higher than is inferred from nucleosynthesis by 50\%, 
although the two are still consistent at a 95\% confidence level. 
It is also interesting to notice that the dominant fraction of energy density 
of the universe is `invisible' (see Table 1). The vacuum energy and CDM mass
occupies over 95\%. Even 3/4 of baryons are invisible. That visible
with optical and X-ray is less than 1\% of the total energy density.

I emphasise that the standard model of the universe involves three 
basic ingredients which are
poorly understood in particle physics: (i) the presence
of small vacuum energy ($\rho_V\simeq 3$ (meV)$^4$), (ii) the presence of
cold dark matter, and (iii) the presence of scalar fields that cause
inflation.
We cannot have successful cosmology without these three substances. 

Although the subjects I have discussed here serve as an interface between
cosmology and particle physics, the particle physics part is still poorly 
understood for most aspects. More successful among others are
speculation of candidate dark matter, and to some extent baryogenesis. 
At least we
have a number of successful models which are related with low energy
phenomenology; yet we cannot choose among these models. The attempt
to understand inflation is much poorer: most models are constructed
without regarding low energy phenomenology or even unified
theories. Furthermore, theorists seem to assume too easily {\it ad hoc}
mechanisms that are not internally motivated in order to
solve `difficulties' the own model create.
Most attempts look like `particle-physics-independent' models.

\section*{Acknowledgements}

I am grateful to W. Hu, A. Jaffe, M. Kawasaki, N. Sugiyama, T. Yanagida 
for discussions.
and to D. Jackson, M. Kawasaki, T. Yanagida and Ed. Turner
for their comments on the manuscript. 
This work is supported in part by the Raymond and Beverly
Sackler Fellowship in Princeton and Grant-in-Aid of the
Ministry of Education of Japan.

\bigskip
\def\hrulefil{\leaders\hrule height0.6pt\hfill\quad}
\def\hrulefill{\leaders\hrule height0.6pt\hfill}
\centerline{TABLE 1}
\centerline{COSMIC ENERGY DENSITY BUDGET}
\hbox to \hsize{\hss\vbox{
\vskip 6pt 
\hrule height 0.6pt
\vskip 2pt
\hrule height 0.6pt
\halign{ \hfil #\hfil & \ \hfil #\hfil & \ \hfil #\hfil & \hfil #\hfil &
\hfil #\hfil \cr
\noalign{\vskip 4pt}
entity & & fraction & &observation \cr
\noalign{\vskip 4pt}
\noalign{\hrule height 0.6pt}
\noalign{\vskip 4pt}
vacuum  & &  70\%  & & invisible \cr
CDM     & &  26\%  & & invisible \cr
baryon  & &   4\%  & &      \cr
        & warm gas    &  &  3\% & invisible \cr
        &{\bf stars~~~} &   &{\bf 0.5\%} & optical \cr
        &{\bf hot gas} & &{\bf 0.5\%} & X-rays \cr
neutrino & &  $>$0.1\% & & invisible \cr
\noalign{\vskip 4pt}
\noalign{\hrule height 0.6pt}
}}\hss}
Note:--bold face means observable components
\bigskip


\end{document}